\documentclass[conference]{IEEEtran}
\usepackage{cite}
\usepackage{url}
\usepackage[pdftex]{graphicx}
\usepackage[]{fancyhdr} % 
\newcommand{\changefont}{\fontsize{9}{9}\selectfont}
\IEEEoverridecommandlockouts
\begin{document}
\title{Inertia Adequacy in Transient Stability Models for Synthetic Electric Grids}
\author{\IEEEauthorblockN{Adam Birchfield}
\IEEEauthorblockA{Dept. of Electrical and Computer Engineering\\Texas A\&M University\\
College Station, TX, USA\\
abirchfield@tamu.edu}
%\thanks{Identify applicable sponsor/s here.  (\textit{sponsors})}
}% <-this % stops a space
\maketitle
\thispagestyle{fancy}
\pagestyle{fancy}
\begin{abstract}
If a disturbance rocks a low-inertia power system, the frequency decline may be too rapid to arrest before it triggers undesirable responses from generators and loads. In the worst case, this instability could lead to blackout and major equipment damage. Electric utilities, to combat this, study inertia adequacy in systems that are particularly vulnerable. This process, involving detailed transient simulations, usually leads to a notion of a system-wide inertia floor. Ongoing questions in this analysis are in how to set the inertial floor and to what extent the location of frequency control resources matters. This paper proposes a new analysis technique that quantifies theoretical locational rate of change of frequency (ROCOF) as a computationally efficient screening algorithm scalable to large systems.  An additional challenge in moving this area forward is the lack of high-quality, public benchmark dynamics cases. This paper presents a synthetic case for such purposes and a methodology for validation, to ensure that it is well suited to inertia adequacy studies to improve electric grid performance.
\end{abstract}
\IEEEpeerreviewmaketitle
\section{Introduction}
Much of the growth in renewable electric power generation in North America and the rest of the world is in technology that does not use large, synchronous machines. Nuclear, hydroelectric, and fossil fuel generation are characterized by massive rotors spinning at synchronous speed, but solar and wind turbines are different. And this difference appears in the way bulk power systems respond to large transient disturbances. When transient power imbalance occurs, the system's near-instantaneous response is for the rotational kinetic energy (inertia) to serve as an immediately-available brief energy resource. Inertia is a buffer to slow frequency decline, until frequency response resources and control systems can adjust generator outputs and load consumption to bring the network back into balance.

The current paper seeks to provide clarity and detail in two ongoing areas of interest in inertia analysis of electric grids. First, the paper presents a fast, conservative bus-level inertia adequacy assessment approach. It is common practice for utilities to assess electric grid inertia capability with system-level calculations that ignore all network effects and compare the amount of inertia online to predetermined conservative inertia floors. Alternatively, utilities do detailed transient stability modeling that would consider network and controls effects; such an approach is quite detailed but can be computationally intensive. This paper's approach is in between. It is very computationally efficient, much faster than a power flow solution, but provides more detail because it considers the locational impacts of where the disturbance occurs, where the inertia sources are located, and where rapid frequency decline would likely cause undesirable effects. The method accomplishes this by calculating the local theoretical rate of change of frequency (ROCOF), which may vary from system-wide ROCOF by significant amounts, particularly in smaller systems, looser-connected systems, and systems with an uneven distribution of inertia.

Second, the work of this paper is predicated on the reality that when it comes to doing research in the area of inertia sufficiency and power system transient dynamics in general, public test data is limited to non-existent. Because of restrictions associated with critical electric infrastructure information (CEII), access to real data is difficult to obtain. Some public test cases do exist, such as the standard 9-Bus WSCC model, but for most public cases the dynamics modeling does not represent the size, scope, and character of today's transitioning grid. This insufficiency is particularly significant with inertia adequacy studies, because the analysis can change with larger systems and how different, separate areas interact. This paper builds on recent work to release a 7000-bus synthetic (fictitious) dataset for general research use. The paper shows how this case has been built to reflect the challenges of modern large-scale bulk electric grid dynamics under the conditions of a rapidly approaching inertia floor.

\subsection{Background on Electric Grid Inertia Evaluation}
The biggest factors affecting how important inertia studies are to any particular electric grid are its overall size and the proportion of generation that comes from synchronous machines. Many large grids worldwide, including those in North America, have been large enough with a large enough proportion of synchronous generators, that historically the connected inertia has been more than sufficient~\cite{denholm2020inertia, tuohy2019implications, agrawal2020study}. Of the three main electric grids in North America, the eastern interconnect (EI) is the largest, most tightly connected, and has the highest proportion of synchronous generators. The western interconnection (Western Electric Coordinating Council, or WECC) is the second largest, with a higher proportion of non-synchronous generation and certain areas less tightly connected to the rest~\cite{agrawal2020study, wecc2021inertia}. Both of these systems are not expected to approach inertia concerns for at least a decade~\cite{denholm2020inertia}.

One example of a system that is beginning to examine inertia in more detail is the Texas interconnection in North America (the Electric Reliability Council of Texas, or ERCOT). ERCOT is the smallest of the three North American grids, and it has been carefully monitoring the large growth in wind generation in the state, particularly in the west, and the potential for low inertia operational conditions to pose a risk for reliable operations~\cite{ercot2022methodologies, sharma2011system}. A recent study has concluded that inertia levels have dropped to a minimum of 130 GW-s in 2017, with detailed analysis indicating that inertia levels above 100 GW-s are sufficient for current ERCOT operating practice~\cite{ercot2018basics}. These studies were done by analyzing the interconnect as a whole, making the typical assumption of a single system-wide frequency.

Another system that is of interest in this area is the synchronous interconnection on the island of Great Britain. Not only is this system smaller in size than ERCOT, and with a high proportion of non-synchronous generation, but there is a closely monitored set of connections between Scotland in the north and England and Wales in the south. These inter-area connections could have the potential to underscore the importance of not only \emph{how much} inertia is available on the system, but also \emph{where} it is located~\cite{tuohy2019implications, HONG20211057, ng2020gridcode,osbouei2019impact}. There is a similar study associated with the grid of Australia~\cite{gu2019review}.

Overall, current literature and industry reports indicate that there are not many active inertia concerns in larger synchronous interconnects, but it is well recognized that this could be a concern in the future. One of the critical questions is how important it is to look at locational inertia, that is, how it is distributed geographically. This area of analysis has received some attention in the last few years, and preliminary work has shown that there can be a substantial difference in the primary frequency response characteristics, critical clearing time, and inter-area oscillations of two grids with identical inertia but different geographical distributions of that inertia~\cite{xu2017investigation, liu2018locational, xu2018location}. Other works have suggested methods to quantify locational inertia and created associated metrics~\cite{xu2018metric,xiao2019framework}. Reference~\cite{tuo2021security} indicates that locational inertia considerations may need to impact unit commitment decisions. Location may also be of importance when it comes to designing frequency response resources such as fast-acting load and energy storage resources acting as virtual inertia~\cite{xu2016application,tuohy2019implications,qureshi2018using}.

\subsection{Background on Dynamics in Synthetic Electric Grids}
In the past few years, great progress has been made in expanding the availability of high-quality, large-scale synthetic datasets for electric grid studies~\cite{birchfield2017synthetic,birchfield2020hicss}. The motivation for building synthetic grids is that actual grid models, due to security concerns, cannot be easily obtained and shared for cross-validating research studies. Synthetic grid creation methodologies use as their input public, geographically-sited actual data on local generation and load. Then the algorithms create a fictitious, realistic transmission network with appropriate engineering parameters to run power flow and other studies. The grids are geographically embedded, validated against realistic data, and have been extended to work on a number of studies including economics and geomagnetic disturbances~\cite{birchfield2016,xu2017econ}.

Full transient stability studies of electric grids requires advanced positive sequence dynamic models representing generator machines and associated control systems. Such cases are critical to the analysis of inertia adequacy and other topics of interest in power system planning. And such cases can be even more difficult to access due to CEII restrictions than the base power flow cases. Some test cases do exist, such as the 9-bus WSCC equivalent case. Preliminary work on developing large-scale synthetic dynamics cases can be found in~\cite{xu2017creation,xu2018modeling}. These references propose a systematic methodology for gathering statistical data on actual power systems dynamics cases, and using those statistics to develop and validate synthetic transient stability models on existing power flow cases. The cases created by these methods extend up to 70,000 buses in size and include multiple model types for synchronous machines, turbine-governor systems, and excitation controllers. The methodology includes validating and tuning the stabilizer controls. This work in building synthetic stability-capable electric grid datasets has opened up continued interest in studying these cases and improving them to apply to specific studies.

\section{Inertia Adequacy Metric Using Localized Theoretical ROCOF}

\subsection{System-wide ROCOF}
The basic relationship between system inertia and the frequency behavior of a power system is found by rearranging the swing equation for a synchronous machine
\begin{equation}
    \dot\omega = \frac{1}{2H} (T_M-T_E)=\frac{-P_{loss}}{2H}
\label{swing}
\end{equation}
where $\dot\omega$ is the time derivative of the machine speed, $H$ is the machine per-unit inertia constant in seconds, $T_M$ and $T_E$ are the per-unit mechanical and electrical torque, and $P_{loss}$ is the power lost during the disturbance, in per-unit on the machine's base.

A zero-order model for the rate of change of frequency (ROCOF) can then be found for the entire system by summing the inertia values for generators, ignoring network effects as if the system were responding as a single generator with a single speed and ROCOF.
\begin{equation}
    ROCOF=\frac{-f_{base,Hz}P_{loss,MW}}{2\sum{H_g S_{g,base}}_g}
\label{system_rocof}
\end{equation}
where $f_{base,Hz}$ is 60 Hz for American systems, $g$ represents each generator with $H_g$ as its per-unit inertia constant and $S_{g,base}$ as its power base in MVA. 

\begin{figure}[th]
\centering
\includegraphics[width=3.5in]{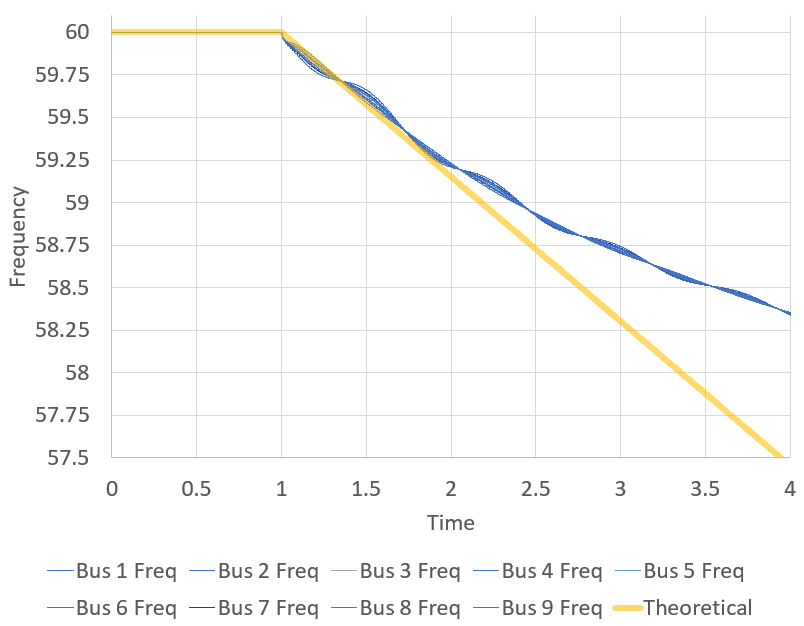}
\caption{Frequency response of 9-bus case to loss of generator three. Yellow trace shows theoretical response calculated by system-wide ROCOF.}
\label{9bus_rocof}
\end{figure}

Take the standard WSCC 9-bus case, for example. Assuming that the contingency involves generator three being outaged, there will be a loss of 85 MW. Generator one has a MVA base of 500 MVA and an inertia constant of 4.728 seconds. Generator two has an MVA base of 250 MVA and an inertia constant of 2.65 seconds. Therefore the expected ROCOF would be -0.8489 $H/s$. In examining the response of Fig.~\ref{9bus_rocof}, the theoretical response is most valid for the initial seconds after the disturbance. Lots of other factors affect frequency response, including
\begin{itemize}
    \item Load response: Constant impedance means that if the voltage drops, power consumption will drop.
    \item Line resistances: As flows change in the system, losses will change and affect the power imbalance.
    \item Speed effects in generator swing equation: As the speed of a machine goes down, it will deviate from the theoretical ROCOF.
    \item Governors and exciters: These control systems are designed to correct the frequency and voltage over time.
    \item Machine damping: The GENROU models used in this example include stator and rotor winding dampings and losses that will include the transient decay of flux stored in the machines' magnetic fields. This effect will change the response from the simple classical swing equation assumed by the system-wide ROCOF equation.
\end{itemize}

ERCOT's inertia monitoring approach is based on the relationship in (\ref{system_rocof}). The designed disturbance is the loss of  2750 MW, and the goal is to not allow the frequency to decline to 59.3 Hz, where the first under-frequency load shedding (UFLS) occurs, before fast frequency response can have time to operate. As a result, ERCOT has determined that the inertia floor, represented by the summation in the denominator of~(\ref{system_rocof}), should be greater than 100 GW-s~\cite{ercot2018basics}. Other systems worldwide have set inertial floors as well, with 135 GW-s for Great Britain, 23 GW-s for Ireland, 125 GW-s for the Nordic interconnect, and 6.2 GW-s for Australia~\cite{tuohy2019implications}.

For this paper, a synthetic 7000-bus grid, known as Texas7000, is applied. The grid is synthetic and does not model the actual ERCOT grid, but it shares a geographic footprint and certain high-level features. It is built using the methods of~\cite{birchfield2017synthetic,birchfield2020hicss} for the transmission network, and~\cite{xu2017creation} for the transient dynamics models. The case, along with the datasets for this paper, are publicly available online at~\cite{tamuelectricgrids}.

\begin{figure}[th]
\centering
\includegraphics[width=3.5in]{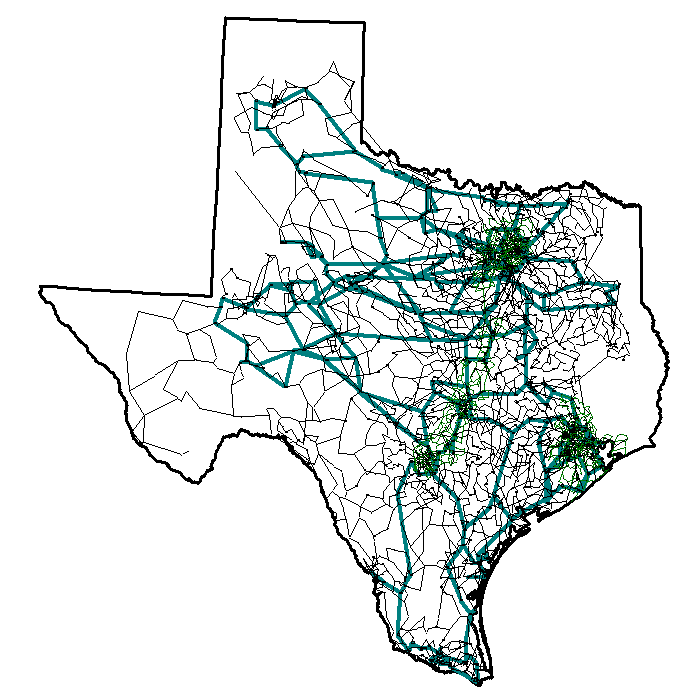}
\caption{Oneline diagram of the synthetic Texas7000 case. This case is synthetic and does not represent any actual grid. The gray lines are 138 kV and 69 kV lines. The teal lines are 345 kV.}
\label{7k_oneline}
\end{figure}

\begin{figure}[th]
\centering
\includegraphics[width=3.5in]{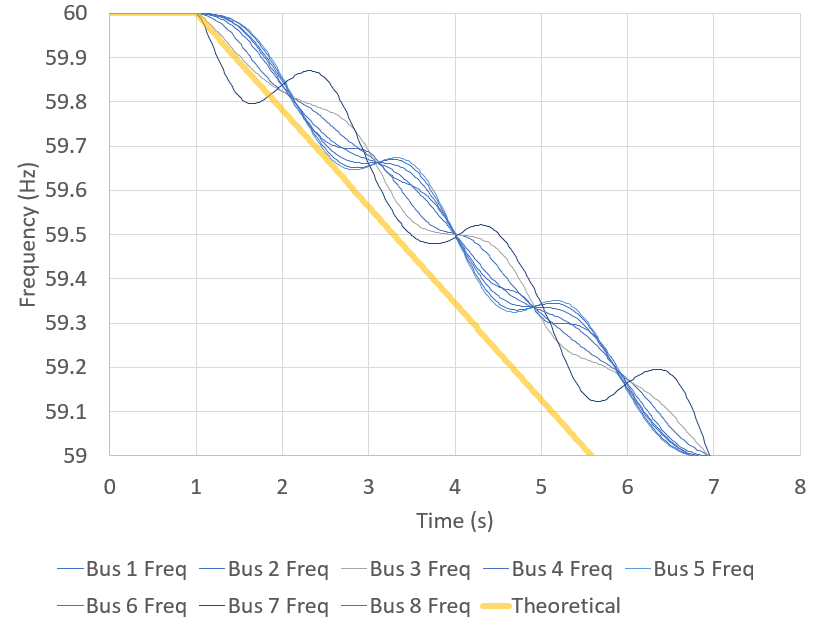}
\caption{Frequency response for selected buses of the Texas7000 case to loss of two large generating units, totaling 2750 MW. Yellow trace shows theoretical response calculated by system-wide ROCOF. Governor and stabilizer responses were removed to isolate inertial response.}
\label{7kbus_rocof_nogov}
\end{figure}

In Fig.~\ref{7kbus_rocof_nogov}, a 2750 MW generator outage is applied, without governor and stabilizer response included, to identify the inertial response. As can be seen, there is close correlation, especially in the first few seconds, with the theoretical system-wide inertia. In general, the system-wide inertia tracks the center of frequency response for the system.

\subsection{Theoretical Locational ROCOF}

As Fig.~\ref{9bus_rocof} and Fig.~\ref{7kbus_rocof_nogov} show, even for a 9-bus system, the frequency varies at least somewhat between the buses. For the larger Texas7000 case, selected buses show significant deviation from one another. Note that this case is a peak planning case, with maximum system inertia online. In cases with lower inertia, and if even more significant disturbances were included, the assumption of a single frequency becomes less valid.

What's important for resilience purposes is not the frequency of the center of the system, but specific frequency as observed at a point in the system, either by a piece of equipment with frequency sensitivity or by a designed under-frequency relay that is measuring local zero crossings. This will determine whether a frequency disturbance will cause undesirable effects. The purpose of the succeeding discussion is to explore a new metric to calculate theoretical ROCOF at each bus in the system following an disturbance.

Synchronous machines are characterized by two main ordinary differential equations: the swing equation (\ref{swing}) and $\dot{\delta}  = \omega$, which shows that the machine torque angle $\delta$ is related to the machine speed $\omega$ through a derivative. Other differential equations apply, showing that the relationship between active power produced by the machine and the electrical torque $T_e$ has dynamics due to the inductances in the rotor and stator. These details are important, but for the sake of calculation we can make the classical assumption that $T_e$ is equal to the instantaneous active power produced by the machine.

Now, at a system level generators are modeled with Norton equivalent currents with the complex matrix equation
\begin{equation}
    \textbf{Y}\cdot V=I
    \label{yvi}
\end{equation}

where $\textbf{Y}$ is the system complex admittance matrix, including loads and non-synchronous generators as shunt impedances, and generators with Norton equivalent shunt admittances corresponding to their direct-axis transient reactance $X^\prime_d$. The complex bus voltage vector $V$ is found by solving the above equation for a given vector of machine current injections $I$. Current injections can be initialized by using the power flow resulting voltages as $V$. Each Norton equivalent for a synchronous generator is related to the state variables as 
\begin{equation}
    I_i=\frac{E^\prime}{X^\prime_d}\angle (\delta_i-\frac{\pi}{2})
    \label{idef}
\end{equation}

and with classical assumptions the electrical torque can be found from $I$ and $V$ to be
\begin{equation}
    T_e=Re[VI^*]
    \label{te_eq}
\end{equation}

The key observation for this paper is that the frequency for any bus is the time derivative of the phase angle of the voltage. And then the ROCOF is the time derivative of the frequency. Hence, taking the second derivative of (\ref{yvi}),
\begin{equation}
    \textbf{Y}\cdot \ddot{V}=\ddot{I}
    \label{yvid2}
\end{equation}

Taking the first and second partial derivatives of (\ref{idef}), we get
\begin{equation}
    \frac{\partial I}{\partial \delta}=\frac{E^\prime}{X^\prime_d}\angle \delta_i
    \label{ideriv1}
\end{equation}
\begin{equation}
    \frac{\partial^2 I}{\partial \delta^2}=\frac{E^\prime}{X^\prime_d}\angle (\delta_i +\frac{\pi}{2})
    \label{ideriv2}
\end{equation}

The total derivatives of $I$ can then be found to be
\begin{equation}
    \dot{I}=\frac{\partial I}{\partial \delta}\cdot \frac{\partial \delta}{\partial t}=\frac{\partial I}{\partial \delta}\omega
    \label{ideriv1t}
\end{equation}
\begin{equation}
    \ddot{I}=\frac{\partial I}{\partial \delta}\dot{\omega}+\omega\cdot (\frac{\partial^2 I}{\partial \delta^2} \omega)
    \label{ideriv2t}
\end{equation}

In the instant after a disturbance, $\omega=0$, which simplifies the equations to $\dot{I}=\dot{V}=0$ and $\ddot{I}=\frac{\partial I}{\partial \delta}\dot{\omega}$, where $\dot{\omega}$ can be found from (\ref{swing}) and (\ref{te_eq}).

In this way, $\ddot{V}=\ddot{V_r}+j\ddot{V_i}$ can be found, which applies to the complex voltage $V=V_r+V_i$. To find the second derivative of the voltage angle $V_a$, use the definition $V_a=\arctan{\frac{V_i}{V_r}}$.

\begin{equation}
    \dot{V_a}=\frac{1}{1+(V_i/V_r)^2} \left(\frac{V_r \dot{V_i}-V_i \dot{V_r}}{V_r^2}\right)=\frac{V_r \dot{V_i}-V_i \dot{V_r}}{V_r^2+V_i^2}
    \label{vap}
\end{equation}
\begin{equation}
    \ddot{V_a}=\frac{V_r \ddot{V_i} -V_i \ddot{V_r}-2 \dot{V_r} \dot{V_i}}{V_r^2+V_i^2}-\frac{2 V_r V_i (\dot{V_i}^2-\dot{V_r}^2)}{(V_r^2+V_i^2)^2}
    \label{vapp}
\end{equation}
which, for $\dot{V}=0$, simplifies to
\begin{equation}
    \ddot{V_a}=\frac{V_r \ddot{V_i} -V_i \ddot{V_r}}{V_r^2+V_i^2}
    \label{vappsimp}
\end{equation}

Thus the process for calculating the theoretical ROCOF at each bus can be summarized as follows:
\begin{enumerate}
    \item Start with initial voltages $V$ from a power flow and $Y$ from system analysis and machine parameters.
    \item Calculate initial $I$, $E^\prime$, $\delta$, $T_e$, and $T_m$ using (\ref{yvi}), (\ref{idef}), and (\ref{te_eq}).
    \item Define disturbance by adjusting $Y$ and $I$ for machine removal or other event.
    \item Get new $V$, $T_e$, $\dot{\omega}$ with (\ref{yvi}), (\ref{te_eq}) and (\ref{swing}).
    \item Calculate $\ddot{I}$ for each generator using (\ref{ideriv2t}).
    \item Use (\ref{yvid2}) to get $\ddot{V}$.
    \item Find the ROCOF (or $\ddot{V_a}$) using (\ref{vappsimp}).
\end{enumerate}

This will result in a vector of ROCOFs, one for each bus. These values can then be compared to the system-wide ROCOF as well as the detailed simulation of $\omega$ and ROCOF.

\section{Designing Synthetic Datasets with Appropriate Inertia Parameters}
Although extensive preliminary work has been done to build synthetic dynamics cases in~\cite{xu2017creation,xu2018modeling}, building synthetic dynamics cases is a difficult problem that can continually be studied to improve the quality and realism of the dynamics cases. In the current study, the prior approach is used as the starting point for the Texas7000 dynamic case, and then modifications are made to increase the realism of the parameters related to inertia and frequency response.

First, the $H$ parameters for the synthetic synchronous generators were set to match a number of validation criteria. In studying actual generator inertia statistics (from literature such as~\cite{ercot2018basics} and from actual dynamics cases), the biggest factors in determining $H$ are the fuel technology and machine size. In this case, three main categories of fuel were considered: nuclear, coal, and gas. Table~\ref{tableGenData} shows the parameters that were then used to form a triangular distribution, guaranteeing the maximum, minimum, and average values for $H$ would be observed. The ranges taper off linearly as the size of the units increase, so that for machines larger than $P_{max}$, $H$ would be equal to the average.

Another important step in setting the $H$ parameters is the important correlation that tends to exist for machines at the same plant. Ordinarily, multiple units at the same plant have very similar designs and therefore similar $H$ constants. Prior synthetic dynamics creation methods did not consider this. For the current case, most generators located at the same substation (with the same fuel type and similar power rating) have the same value for $H$.

\begin{table}[!t]
\renewcommand{\arraystretch}{1.3}
\centering
\caption{Generator probability distribution parameters by fuel type, aggregated from literature and industry cases.}
\label{tableGenData}
\begin{tabular}{|c|c|c|c|c|}
\hline
Fuel & $H_{max}$ & $H_{min}$ & $H_{avg}$ & $P_{max}$\\
\hline
\parbox[t]{.85in}{Nuclear}&5.2&3.8&4.2&10000\\
\hline
\parbox[t]{.85in}{Coal}&6&2&3.2&3000\\
\hline
\parbox[t]{.85in}{Gas}&10&1&4.3&2000\\
\hline
\end{tabular}
\end{table}

\begin{figure}[th]
\centering
\includegraphics[width=3.5in]{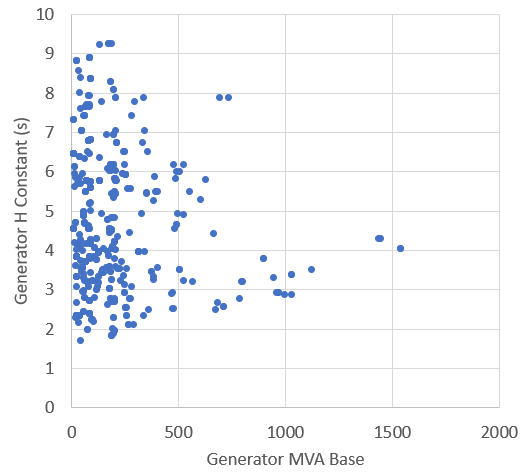}
\caption{Distribution of generator H constants in synthetic Texas7000 case. By design, the variation is greater for smaller units, with the averages still matching what is given in the table.}
\label{H_dist}
\end{figure}

For this example test case, the total inertia turns out to be 402 GW-s, which is in line with reported values from ERCOT~\cite{ercot2018basics}. The distribution of generator inertia parameters can be seen in Fig.~\ref{H_dist}.

The other main aspect of creating synthetic stability cases is to define where in the system various impacts of frequency decline would be detected. In this case, the ERCOT under frequency load shedding (UFLS) structure was used, with loads probabilistically assigned to each of the three levels of load shedding: 5\% at 59.3 Hz, 10\% at 58.9 Hz, and 10\% more at 58.5 Hz. It is important to model the fact that the loads will trip if the frequency \emph{at that location} goes below the frequency target. Additional loads can be assigned as fast frequency response resources, which will trip (desirably) after a voltage sag below 59.7 Hz that lasts more than 25 cycles. Each generator should also be monitored as to what the frequency is doing in that location, in case there is a risk of generator damage or tripping due to local frequency conditions.

The resulting synthetic case and all dynamics data has been made available at~\cite{tamuelectricgrids}.

\section{Results from Example Application to Inertia Adequacy Study}
In this section, an inertia adequacy analysis is done to the synthetic Texas7000 case, comparing the methodologies of system-level inertia floors, bus-level theoretical ROCOF, and full transient stability simulations.

\subsection{Creating Scenario Bank to Run}
The inertia impacts depend on which scenario is studied. A scenario is defined by two main things. One, the base loading level and dispatch, including which generators are online. And two, the transient contingency that occurs. For system-level inertia analysis, the contingency doesn't depend on which units are outaged, just the total MW lost. But the location of the generator tripping can have an impact on the frequency response.

For this study, 163 different contingencies are considered by probabilistically sampling generator outages. All contingencies are considered large outages, with greater than 800 MW of generation lost. Multiple unit outages are considered for a single generating plant. Most of the contingencies are sized below 2750 MW, as the standard largest contingency. But to explore the extreme cases, a few of the contingencies involve multiple units from different sites outaging, with much larger total outages. 

These contingencies are to be analyzed on 125 different loading and dispatch levels. The loading was changed to represent the most extreme cases from 15,000 MW of load up to peak load of 75,000 MW. In general, the lower loading cases are suspected to be more susceptible to frequency issues due to lower inertia. Each scenario also has a different level of wind generation, from 10,000 MW up to 30,000 MW, which is the maximum wind level in the case. Future work could also look at higher levels of wind added on top of the existing case. In each scenario, a unit commitment and optimal power flow are performed to determine which units are online and what amount of power they are producing. Wind and nuclear are dispatched to full output, to the extent allowed by transmission constraints. Combining with the 163 contingencies to consider, in total a set of about 20,000 scenarios were created for analysis. 

\subsection{Inertia Results}
\begin{figure}[th]
\centering
\includegraphics[width=3.5in]{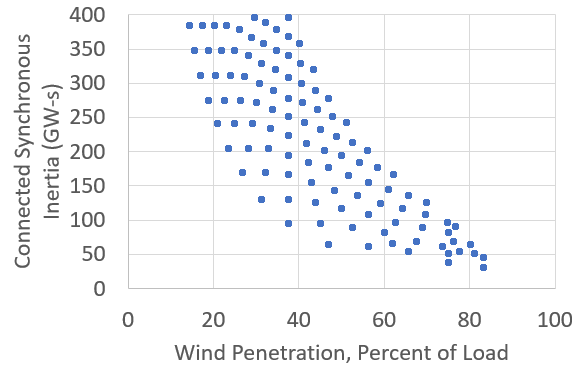}
\caption{Loading cases studied, inertia levels compared to the percent of generation contributed by wind.}
\label{Scenario_Wind}
\end{figure}

\begin{figure}[th]
\centering
\includegraphics[width=3.5in]{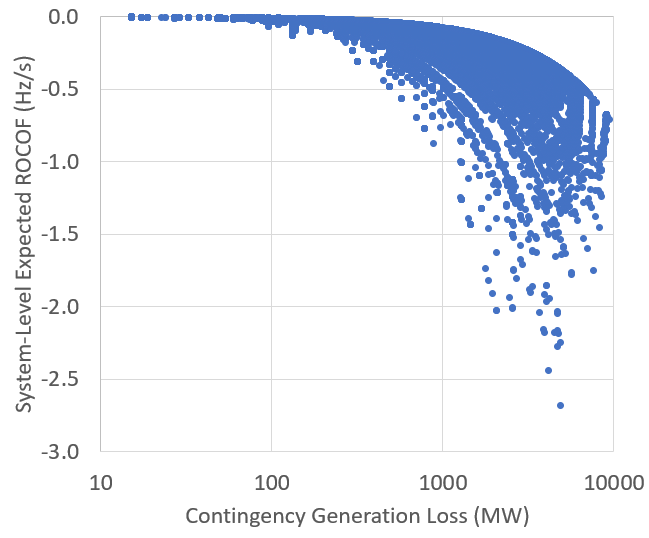}
\caption{Scatter plot of generator loss scenarios, showing the theoretical system-level ROCOF compared to the magnitude of generation lost.}
\label{All_scenario_loss}
\end{figure}

\begin{figure}[th]
\centering
\includegraphics[width=3.5in]{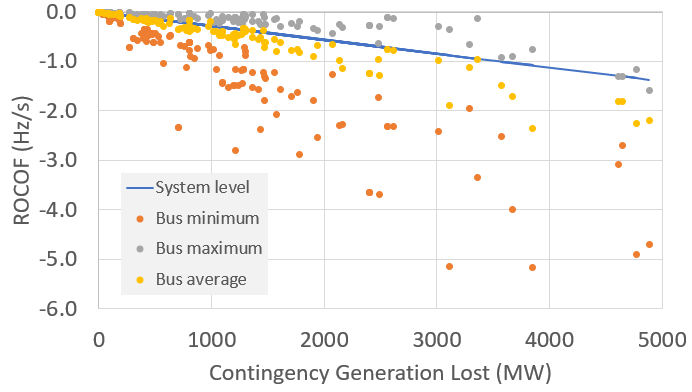}
\caption{Plot of ROCOF calculations for all the contingencies in the scenario with 30 GW of load, 18 GW of wind.}
\label{ROCOF_30_18}
\end{figure}

\begin{figure}[th]
\centering
\includegraphics[width=3.5in]{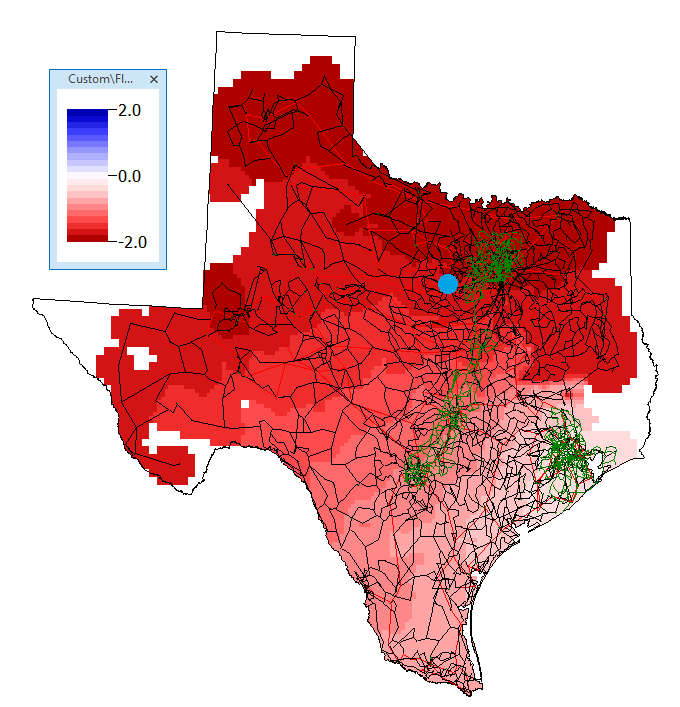}
\caption{Contour of the bus-level theoretical ROCOF for one example case. Location of the disturbance shown with blue dot.}
\label{Contour_ROCOF}
\end{figure}

\begin{figure}[th]
\centering
\includegraphics[width=3.5in]{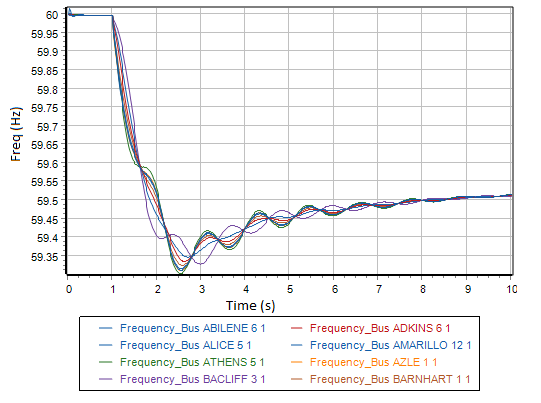}
\caption{Plot of bus frequencies for example case under full transient simulation.}
\label{Freq_plot}
\end{figure}

The results of this analysis are shown in Figs.~\ref{Scenario_Wind}--\ref{Freq_plot}. In Fig.~\ref{Scenario_Wind}, each of the 125 loading cases are shown, comparing the wind penetration percentage to the total connected synchronous inertia. Comparing to the similar chart in \cite{ercot2018basics}, it is clear that this study includes many cases beyond what ERCOT in particular has experienced before. The record is about 55\% of the generation coming from renewable sources, and the reported lowest inertia experienced is 130 GW-s.

In Fig.~\ref{All_scenario_loss}, all the 20,000 scenarios (combination of loading case and contingency event) are plotted with the quick analysis of the system-level expected ROCOF, compared to the amount of generation outaged by the scenario's contingency. This plot shows that many contingencies were considered beyond the 2750 MW design contingency. A value of ROCOF below about -0.5 Hz/s could be potentially of concern because UFLS could be triggered in about 84 cycles, which may be faster than frequency response resources could detect and remedy the problem. 

For each of the 20,000 scenarios, a locational theoretical ROCOF analysis was also done in accordance with II.B. So for each scatter dot in~\ref{All_scenario_loss}, there is also an array of bus-level ROCOF values. A single loading case is chosen for the data shown in Fig.~\ref{ROCOF_30_18}, to give an example. This is the scenario with 30 GW of load and 18 GW of wind generation (around the boundary of realistic cases for today's grid). Here, the range of ROCOF values for each case (maximum, minimum, and mean) is compared with the system-level theoretical value (linear function of contingency disturbance in (\ref{system_rocof})). It's clear in this analysis that the average tracks reasonably well with what the zero-order expectation of the overall frequency predicts. But some buses, at least briefly, experience a much larger drop in frequency than average.

Zooming in on one particular scenario provides even greater clarity. In this case, it is a 1300 MW generator outage selected because it showed some buses with very low ROCOFs in the bus-level analysis. Contouring these geographically in Fig.~\ref{Contour_ROCOF}, it seems that the northern portion of the grid, above the disturbance location, experience a greater drop than would be expected by the overall size of the disturbance (it is, after all, only half the size of the design contingency). 

Running this study in a commercial simulator with full transient dynamics modeling, the time series of eight selected buses is shown in Fig.~\ref{Freq_plot}. The differences in frequency can be seen, and the frequency nadir is low but would not ordinarily cause UFLS. But in this case, UFLS was indeed triggered, at five substations in the northern and eastern parts of the grid. The quick recovery around 2.5 seconds is in part due to UFLS that occurred. Most of the UFLS were not triggered, because most of the grid did not see a low enough frequency. This case illustrates the potential for this analysis to provide greater insight than the system-wide metric can.

\subsection{Computational Speed Analysis}
The system-level inertia analysis is extremely fast, since it just sets an inertia floor (depending on maximum tolerable ROCOF) for a design contingency. Any contingency with a lower loss than the design, and any scenario with lower inertia, is considered to require further study. While this is the fastest study, it does have the potential to miss localized effects, and to be overly conservative.

Full transient stability modeling involves numerically integrating ordinary differential equations over 10 seconds or so, then assessing what happened to the frequency using signal processing methods. This is by far the most detailed analysis, but it does depend on accuracy of individual models, may require significant human intervention in case certain scenarios have difficulty solving, and may take 1-3 minutes to solve each for a system the size of Texas7000.

The new locational theoretical ROCOF, as part of an inertia adequacy study, is extremely fast. A single scenario can be evaluated with just two sparse matrix solutions, much faster than a power flow (milliseconds). For a single loading scenario, in 1-2 minutes it could test thousands of contingencies. And in this case 20,000 different loading and contingency scenarios were tested in about 3 hours on a single laptop. Hence this method provides more detail for a fast and scalable solution.

\section{Summary and Future Work}
An important technical need in moving toward higher levels of wind or solar generation in bulk power grids is resilience to large-scale generator outages in the form of frequency response. The key initial frequency response characteristics are the synchronous inertia of large generators. As the grid changes, it is important to monitor inertia and the capabilities of grid dynamics. In this paper, a methodology for building synthetic dynamics cases with appropriate inertia characteristics is presented. It is assessed using both system-wide inertia calculations and full transient dynamics. In addition, a new metric is presented which can very quickly identify the local theoretical ROCOF at each bus to identify if there are regional frequency control challenges.

Moving forward, continued development of high-quality, large-scale synthetic dynamics cases will open up many opportunities for research and development in this area. It is important for researchers to be able to cross-validate on public cases. In addition, the application of local theoretical ROCOF to other sources of inertia and frequency response, such as governor settings, storage acting in frequency response, and synchronous condensers, should be considered.

\bibliographystyle{IEEEtran}
\bibliography{IEEEabrv,birchfield_references.bib}

\end{document}